\input{aipcheck}

\documentclass[
 final            
,numberedheadings 
  ]
  {aipproc}

\layoutstyle{8x11double}

\begin{document}

\title{Lunar Imaging and Ionospheric Calibration for the Lunar Cherenkov Technique}


\author{R. McFadden}{
  address={\emph{formerly} Astron, Oude Hoogeveensedijk 4, 7991 PD, Dwingeloo, The Netherlands.}
}

\author{O. Scholten}{
  address={Kernfysisch Versneller Instituut, University of Groningen, 9747 AA, Groningen, The Netherlands.}
}

\author{M. Mevius}{
  address={Kernfysisch Versneller Instituut, University of Groningen, 9747 AA, Groningen, The Netherlands.}
}

\begin{abstract}
 The Lunar Cherenkov technique is a promising method for UHE neutrino and cosmic ray detection which aims to detect nanosecond radio pulses produced during particle interactions in the Lunar regolith. For low frequency experiments, such as NuMoon, the frequency dependent dispersive effect of the ionosphere is an important experimental concern as it reduces the pulse amplitude and subsequent chances of detection. We are continuing to investigate a new method to calibrate the dispersive effect of the ionosphere on lunar Cherenkov pulses via Faraday rotation measurements of the Moon's polarised emission combined with geomagnetic field models. We also extend this work to include radio imaging of the Lunar surface, which provides information on the physical and chemical properties of the lunar surface that may affect experimental strategies for the lunar Cherenkov technique. 
 \end{abstract}

\maketitle


\section{Introduction}

Lunar Cherenkov experiments employ ground-based radio telescopes to search for nanosecond radio pulses produced through the Askaryan effect. This effect has been confirmed in a variety of media including silica to approximate the lunar regolith \cite{Saltzberg2001} . Due to the limitations of terrestrial accelerators, these experiments can not be performed in the ultra-high energy (UHE) range, however, the Cherenkov emission spectra for UHE particle interactions have been described analytically in a number of publications (for e.g. \cite{Alvarez2006}) and the resulting pulse properties can be related to medium-dependant parameters through these spectra. UHE particle detection using the lunar Cherenkov technique is essentially a pulse detection problem and an accurate knowledge of the expected pulse characteristics can be used to optimise experimental strategy and increase the received pulse amplitude. Here, we explore how lunar surface properties affect pulse production at the lunar surface and how knowledge of these properties may also be used to improve ionospheric calibration techniques. We review how lunar surface properties have historically been studied through the radio astronomy techniques of profiling, visibility fitting and imaging and provide some initial calculations for imaging at low radio frequencies. 

\section{Pulse Characteristics}

A high energy particle interacting in the lunar dielectric produces a cascade of secondary particles with a charge asymmetry due to Compton scattering and positron annihilation. This results in coherent Cherenkov emission as the time-varying charge excess moves through the dense lunar medium. The angular diffraction pattern of the emission peaks at the Cherenkov angle where emission along the length of the shower is compressed and the pulse width is determined only by the lateral dimension of the shower \cite{Alvarez2006}. Away from the cone, decoherence effects along the length of the shower contribute to a widening pulse and a loss of high frequency components in the spectrum. High frequency experiments are therefore limited in their detection geometries but have a lower energy detection threshold (geometrically favourable events peak at a few GHz). At lower frequencies, where the wavelength is comparable to the shower length, the emission is nearly isotropic. Experiments at these frequencies are less sensitive to interaction geometry and have a larger detection volume. However, this is at the cost of a higher energy detection threshold. 

A good understanding of the expected emission and pulse properties can drive experimental strategy and increase the chances of detection. However, while the expected emission has been parameterised, there are still some contributing factors which are poorly understood (for a good summary see \cite{James2012unanswered}). These factors include surface roughness which, depending on its spatial scale, may destroy the coherence of the emission or bias events toward unfavourable geometries, and the radio-absorbency depth profile of the regolith and sub-regolith layers, all of which affect the detector volume. Simulations are also sensitive to uncertainties in lunar dielectric properties and surface composition as important characteristics of the emission spectra such as the Cherenkov angle, turnover frequency and the angular diffraction width are determined by the electrical properties of the interaction medium. 

Once a pulse has left the lunar surface, more uncertainties are introduced into the pulse characteristics during propagation. Propagation through the Earth's ionosphere introduces dispersion which reduces coherency within the pulse and subsequent chances of detection. Full pulse amplitude can be recovered using matched-filtering techniques and this requires an accurate knowledge of the dispersion characteristic which is parameterised by the instantaneous Total Electron Content (TEC) of the ionosphere. Our main motivation for studying the lunar surface is to understand the Moon's polarisation properties so that the dispersive effect of the ionosphere can be calibrated via lunar Faraday rotation measurements combined with geomagnetic field models. We also extend this work to include radio imaging of the lunar surface, which may provide further information on the physical and electrical properties of the lunar surface relevant to the lunar Cherenkov technique. 

\section{Lunar Astronomy}

The observed lunar intensity and polarisation are related to the thermal and electrical properties of the lunar surface \cite{Heiles1963}. The Moon can be considered as a smooth, homogenous spherical planet with a power reflection coefficient (and hence emissivity) which is a function of the dielectric constant of the surface material and the angle between the direction of propagation and the surface normal. Deviations from an ideal, smooth sphere can be caused by surface roughness, changes in the lunar surface composition and variations of temperature, density or electrical properties of the lunar material with depth. 

Lunar intensity and polarisation distributions exhibit a strong dependence on the radial axis which favoured lunar profile scanning for early investigations of the Moon. Since then, three main methods for obtaining information about lunar emission via radio astronomy techniques have emerged: lunar profile scanning, visibility data analysis and interferometric imaging. 

\subsection{Model Fitting}

The strong radial dependence and angular symmetry of lunar emission \cite{Heiles1963} has made lunar profile and visibility fitting an attractive method for investigating the lunar surface using radio astronomy techniques. A lunar profile is obtained by taking a scan each in right ascension and declination through the centre of the Moon to obtain a profile of the intensity or polarisation distribution across the disk.   

The lunar intensity profile shows limb darkening due to surface roughness effects \cite{Davies1966} while the intrinsic thermal radiation of a planetary object appears increasingly polarised toward the limb, when viewed from a distant point such as Earth \cite{Heiles1963}. The polarised emission is radially aligned due to the changing angle of the planetary surface toward the limb combined with Brewster angle effects. This effect is known as `Brewster angle highlights' or limb brightening and is sensitive to changes in the dielectric constant through the power reflection coefficient. An increase in the dielectric constant increases the degree of polarisation (ratio of polarisation to brightness temperatures) as the difference between the parallel and perpendicular Fresnel reflection coefficients is increased. Surface roughness effects will also modify the polarisation response, therefore the observed polarisation is related to the effective dielectric constant which is a mix of the true dielectric constant and surface roughness effects \cite{Schloerb1976} \cite{Heiles1963}.   

Properties of lunar sub-surface material may also be inferred from lunar profile analysis. Measurements taken at longer wavelengths will probe greater depths of material and help to build a depth profile of the sub-surface thermal and electrical properties. An increase in the observed dielectric constant with wavelength has been attributed to a higher density of lunar material at greater depths \cite{Davies1966}, while attempts to determine the subsurface heat flow may also reveal information about the sub-surface layers as temperature gradient estimates (based on limb-darkening models) conflict with those measured at the Apollo site suggesting either inhomogeneities with depth or a solid rock layer at about 10m \cite{Schloerb1976}.       

Alternatively, asymmetries in the Moon's brightness distribution can be measured in the visibility function which is the response of an interferometer antenna pair or baseline. The visibility function samples the two-dimensional Fourier transform of the spatial brightness distribution, however, it reduces to a one-dimensional Hankel transform for a circularly symmetric object such as the Moon. The visibility function is also sensitive to the value of the dielectric constant and surface roughness effects in both the polarisation distribution \cite{Heiles1963} and intensity limb darkening \cite{Schloerb1976} .      
 
Radar measurements and analysis of samples from Apollo missions provide complementary measurement techniques, however, historically there has been some disagreement between the measurements obtained from these techniques and radio astronomy data. These differences have been attributed to inhomogeneities in the electrical properties of the lunar surface and different wavelengths probing different depths of near-surface layers (radar wavelength relates differently to observation depth and this relation is sensitive to grazing angle for any instrument) \cite{Davies1966}. The possibility of inhomogeneities in the lunar surface makes mapping more important, particularly at a variety of wavelengths, and observing the lunar surface through radio telescopes offers the opportunity to observe the lunar emission with the same geometry, propagation effects, calibration and instrumentation issues as during the lunar Cherenkov experiments.
 
\subsection{Imaging}

There are few examples of lunar mapping in the literature. In 1965 Heiles and Drake \cite{Heiles1963} used the NRAO 300-ft telescope to produce a map of the lunar polarised emission at 1.4 GHz. They used a 'wobbling' technique (i. e. scanning back and forth across the Moon in declination) and produced a contour map of the difference in brightness temperature measured by two cross-polarised feeds. They found the Moon to be very smooth -- in agreement with radar maps -- but noticed asymmetry in the peaks of the polarisation most likely caused by surface inhomogeneities. In particular, some mare regions had reduced polarisation indicating either an increase in surface roughness or a decrease in dielectric constant. In 1972 Moffat \cite{Moffat1972} produced polarimetric maps, also at 1.4 GHz, using aperture synthesis techniques. He found that the Moon is well described by a simple model except for local deviations in fractional polarisation and intensity brightening toward the limb. These polarisation deviations were not consistent with uniform dielectric constant and surface roughness although the noise level in the map was considered too high to draw any definite conclusions. More recently, in 2002,  Poppi et. al. \cite{SPORT2002} investigated the Moon as a polarised calibrator at centimetre and millimetre wavelengths. They produced maps of the polarised lunar emission at 8.3 GHz with the Medicina 32-m radio telescope. They found that the the polarised emission was not homogenous and noted a decrease in the expected linearly polarised signal which was attributed to surface roughness effects and antenna pattern averaging. An excess of the Stokes $U$ parameter in integrated polarisation also suggested inhomogeneities of the lunar regolith physical properties on a large scale. In each case, polarisation defects were observed which may be related to lunar surface features or regions of varying thermal and electrical properties with further lunar mapping.

\subsection{Using Lunar Observations for Ionospheric Calibration}

A good understanding of the expected thermal emission from the lunar surface can also help to calibrate the effects of pulse propagation. In particular the frequency-dependent refractive index of the Earth's ionospheric plasma causes a differential additive delay across the bandwidth of a propagating pulse which results in a loss of coherency and reduction of the received pulse amplitude. This effect is most dramatic for low frequency experiments such as the NuMoon experiments which operate in the 100-200 MHz range. A reduction of the received pulse amplitude not only affects the chances of pulse detection but also the neutrino energy threshold. The amplitude of the pulse at the lunar surface is related to the charge excess in the particle shower which is roughly proportional to the number of particles in the electromagnetic cascade which in turn is proportional to the energy of the shower and the energy of the original neutrino. Therefore as the minimum detectable pulse height is increased, the minimum detectable neutrino energy is also increased.

Pulse amplitude can be recovered using coherent dedispersion techniques which require knowledge of the real-time ionospheric conditions. Current Ionospheric TEC products available online include Global Positioning System (GPS) satellite data and ground-based ionosonde data. Ionosondes probe the peak transmission frequency (fo) through the F2-layer of the ionosphere which is related to the ionospheric TEC, however, there are known inaccuracies in TEC estimates obtained from ionosonde measurements as they are empirically derived \cite{Mcfadden2011}.  TEC measurements can also be derived from dual-frequency GPS signals and maps of this data are available online from several sources \cite{Nasa} \cite{IPS}. These maps contain estimates derived from GPS measurements processed with Kalman filters and combined with the IRI-2007 \cite{IRI2007} ionospheric model which is driven by real-time foF2 observations from ionosondes \cite{IPS}. They display a total delay from GPS satellites (at an altitude of $\sim$ 20000 km) to the ground, mapped to a vertical path. Mapping includes both ionospheric and plasmaspheric components, modelled separately, and combined to produce an estimate of total delay. The ionosphere can be modelled as a single layer model which assumes all free electrons are concentrated in an infinitesimally thin shell and removes the need for integration through the ionosphere. This is scaled according to the standard Gallagher model to account for the plasmaspheric contribution to the TEC \cite{Gallagher1988}. While the time resolution and availability of ionospheric products has improved a lot in the last few years, there are still inaccuracies associated with using these products to calibrate the ionosphere for the lunar Cherenkov technique. In particular there are significant errors associated with mapping to and from vertical paths. New methods of ionospheric monitoring are required particularly for the current low-frequency lunar Cherenkov experiments and, as the solar cycle enters a more active phase, accurate pulse dedispersion will become a more important experimental concern at all frequencies.

We are continuing to investigate a new technique to obtain TEC measurements that are both instantaneous and line-of-sight to the Moon \cite{Mcfadden2011}. Ionospheric TEC can be deduced from the Faraday rotation measurements of a polarised source combined with a geomagnetic field model, which are more stable than ionospheric models (IGRF magnetic field values are accurate to better than 0.01\% \cite{IGRF}). We propose to use this method with lunar thermal emission as the polarised source.

Initial investigations of this technique were performed with the LUNASKA collaboration using the 22-m telescopes of the Australia Telescope Compact Array with a center frequency of 1384 MHz \cite{Mcfadden2011}. At these frequencies, planetary synthesis imaging and polarimetry require a complete set of antenna spacings and enough observing time for earth rotation synthesis. Faraday rotation measurements obtained through synthesis imaging will therefore be averaged over the entire observational period and not contain any information on short-term ionospheric structure. Depending on the chosen experimental strategy, the lunar imaging baseline requirements may also conflict with the unique constraints of a lunar Cherenkov experiment. For UHE particle detection, long antenna spacings may be preferred to minimise the level of lunar noise correlation between antennas or short spacings may be used to minimise the relative geometric delays between antennas. To overcome these limitations we developed a method of obtaining lunar Faraday rotation estimates in the visibility domain which makes use of angular symmetry in the lunar polarisation distribution. Preliminary comparison of this technique to GPS data show that both data sets exhibit similar features which can be attributed to ionospheric events, however, more observations are required to investigate this technique further. Ionospheric dispersion effects are more significant for low frequencies experiements such as Numoon which is currently using the Low Frequency Array (LOFAR) for UHE neutrino detection experiements  \cite{Singh2011}. LOFAR's increased sensitivity and more complicated baseline geometry (compared to the ATCA's east-west configuration) make imaging techniques more suitable than visibility function analysis for lunar studies.  

\subsection{Initial Calculations for Low Frequency Lunar Imaging}

We first want to map the low frequency lunar emission to determine brightness temperature in different regions and any correlation to surface features such as mare or highland regions. Mapping at low frequencies will probe a different surface scale and depth to the existing lunar maps. The LOFAR telescope operates low band antennas (LBA) at 10-80 MHz and high band antennas (HBA) at 120-240 MHz. At these frequencies, the sky brightness temperature is dominated by galactic radiation which has a strong frequency dependence given by, $T_{\mathrm{sky}} = T_{\mathrm{s0}}\lambda^{2.55}$ where $T_{\mathrm{s0}} = 60 \pm 20$ K for galactic latitudes between 10 and 90 degrees. Assuming a lunar temperature of 241 K, there is a turnover frequency at approximately 174 MHz where the Moon is the same temperature as the background sky and can not be detected. Below this frequency, the Moon is a negative source or sink and above, the Moon becomes an increasingly strong positive source. Choice of observation frequency in the LOFAR bands is therefore non-trivial.   

\begin{figure}[htbp]
  \includegraphics[height=.27\textheight]{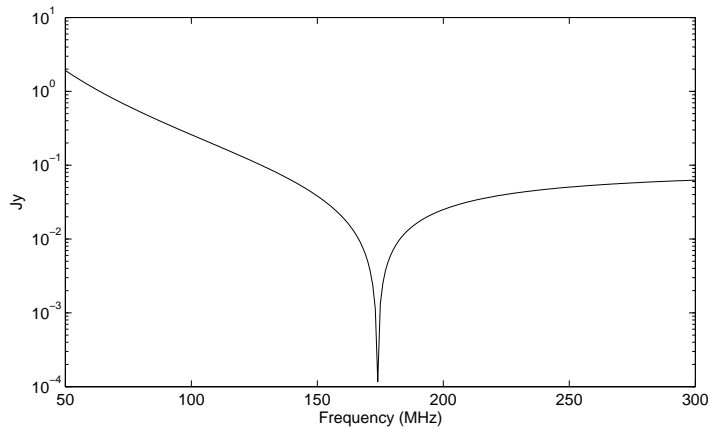}
  \caption{Lunar flux density relative to the sky temperature, per LOFAR core station pixel beam. The minima occurs when the Moon is the same temperature as the background sky.}\label{flux}
\end{figure}

\begin{figure}
  \includegraphics[width=.35\textwidth]{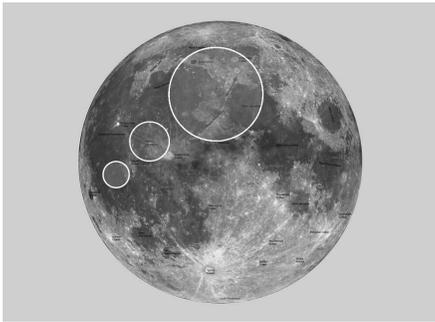}
  \caption{Pixel beams of the LOFAR core stations superimposed onto an image of the Moon. Beams (in increasing size) correspond to 150, 100  and 50 MHz.}\label{resolution}
\end{figure}

The effective lunar flux density (see Figure \ref{flux}) was calculated for the LOFAR observing bands using just the core LOFAR stations, which correspond to a maximum baseline of approximately 2 kilometres. The lunar flux is stronger at lower frequencies, however, LOFAR has significantly better sensitivity in the HBA band. Therefore there is an optimal window to observe the Moon at the lower end of the HBA band (i. e. $\approx$ 110 MHz). At this frequency, the resolution for a maximum baseline of 2 kilometres is 3.5' (see Figure \ref{resolution}) and the Moon is a sink with an expected flux density of  $-$.2 Jy per pixel beam. The instrumental sensitivity is inversely proportional to the square root of observing time-bandwidth product. For a one hour, one sub-band (for technical reasons, this is $\approx$ .2 MHz) observation, a sensitivity equivalent to $\approx$ 30$^\circ$ K is possible. To improve this to $\approx$ 3$^\circ$ K requires a factor of 100 in time and frequency for example, 10 hours observation and 10 sub-bands ($\approx$ 2 MHz). 

Extension to polarimetric imaging introduces some additional complications. Polarisation at the limb of the Moon is only an 8\% effect \cite{Heiles1963} and this is further reduced by beam smearing. The Medicina observations had a resolution of 4.8' (FWHM) and the observed polarisation was consequently reduced to approximately 5\%. The LOFAR resolution of 3.5' is only slightly more optimistic, therefore we can anticipate to lose a factor of $\approx$ 20 (with respect to the imaging case) due to incomplete polarisation. Detailed polarimetric maps at low frequencies therefore require careful calibration and improved imaging techniques to achieve the sensitivity required for mapping surface features, however, it is possible to attempt ionospheric calibration without producing full lunar maps. Lunar polarisation is radially aligned around the limb of the Moon, therefore limb pixels can be stacked to improve the sensitivity provided that each pixel is rotated according to its position on the limb so that the radial polarisation vectors would be aligned. Any deviations from the modelled polarisation can then be attributed to Faraday rotation and converted to ionospheric TEC measurements.  At 110 MHz, approximately 30 beams can be fit around the limb of the Moon offering a gain of 30. Measurements could be also taken at different frequencies so that Faraday rotation measurements can further be constrained across the frequency range offering additional gain. Due to phase screening effects, the polarised background is less well understood therefore this will also have to be carefully considered.

\section{Further Work}

Further radio astronomical observations of the lunar surface may assist UHE neutrino detection using the lunar Cherenkov technique by revealing more about the lunar surface properties which affect pulse production and propagation. We are continuing our efforts to optimise low frequency lunar imaging and calibration techniques with a view to producing a lunar intensity image, determining Faraday rotation to assist ionospheric calibration and to explore the polarisation defects previously observed with radio telescopes.

\bibliographystyle{aipproc}   

\bibliography{My_bib2}

\begin{thebibliography}{15}
\expandafter\ifx\csname natexlab\endcsname\relax\def\natexlab#1{#1}\fi
\providecommand{\enquote}[1]{``#1''}
\expandafter\ifx\csname url\endcsname\relax
  \def\url#1{\texttt{#1}}\fi
\expandafter\ifx\csname urlprefix\endcsname\relax\def\urlprefix{URL }\fi
\providecommand{\eprint}[2][]{\url{#2}}

\bibitem[{Saltzberg} et~al.(2001)]{Saltzberg2001}
D.~{Saltzberg}, P.~{Gorham}, D.~{Walz}, C.~{Field}, R.~{Iverson}, A.~{Odian},
  G.~{Resch}, P.~{Schoessow}, and D.~{Williams}, \emph{Phys. Rev. Lett.}
  \textbf{86}, 2802 (2001).

\bibitem[{Alvarez-Mu{\~n}iz} et~al.(2006)]{Alvarez2006}
J.~{Alvarez-Mu{\~n}iz}, E.~{Marqu{\'e}s}, R.~A. {V{\'a}zquez}, and E.~{Zas},
  \emph{Phys. Rev. D} \textbf{74}, 023007 (2006).

\bibitem[{James}(2012)]{James2012unanswered}
C.~W. {James}, \emph{Nuclear Instruments and Methods in Physics Research A}
  \textbf{662}, 12 (2012).

\bibitem[Heiles and Drake(1963)]{Heiles1963}
C.~E. Heiles, and F.~D. Drake, \emph{Icarus} \textbf{2}, 281 (1963).

\bibitem[{Davies} and {Gardner}(1966)]{Davies1966}
R.~D. {Davies}, and F.~F. {Gardner}, \emph{Australian Journal of Physics}
  \textbf{19}, 823 (1966).

\bibitem[{Schloerb} et~al.(1976)]{Schloerb1976}
F.~P. {Schloerb}, D.~O. {Muhleman}, and G.~L. {Berge}, \emph{Icarus}
  \textbf{29}, 329--341 (1976).

\bibitem[{Moffat}(1972)]{Moffat1972}
P.~H. {Moffat}, \emph{MNRAS} \textbf{160}, 139--154 (1972).

\bibitem[{Poppi} et~al.(2002)]{SPORT2002}
S.~{Poppi}, E.~{Carretti}, S.~{Cortiglioni}, V.~D. {Krotikov}, and E.~N.
  {Vinyajkin}, \enquote{{The Moon as a calibrator of linearly polarized radio
  emission for the SPOrt project},} in \emph{Astrophysical Polarized
  Backgrounds}, 2002, vol. 609 of \emph{AIPC}, p. 187.

\bibitem[{McFadden} et~al.(2011)]{Mcfadden2011}
R.~{McFadden}, R.~D. {Ekers}, and J.~D. {Bray}, \enquote{{Ionospheric
  propagation effects for UHE neutrino detection using the lunar Cherenkov
  technique},} in \emph{ICRC}, 2011, vol.~4 of \emph{ICRC}, p. 280.

\bibitem[{NASA JPL}(2012)]{Nasa}
{NASA JPL}, {Ionospheric TEC Map},
  \url{iono.jpl.nasa.gov//latest_rti_global.html} (2012).

\bibitem[{Ionospheric Prediction Service}(2012)]{IPS}
{Ionospheric Prediction Service}, {Australasia Total Electron Content.},
  \url{www.ips.gov.au/Satellite/2/1} (2012).

\bibitem[Bilitza and Reinisch(2008)]{IRI2007}
D.~Bilitza, and B.~Reinisch, \emph{J. Adv. Space Res.} \textbf{42}, 599--609
  (2008).

\bibitem[{Gallagher} et~al.(1988)]{Gallagher1988}
D.~L. {Gallagher}, P.~D. {Craven}, and R.~H. {Comfort}, \emph{Advances in Space
  Research} \textbf{8}, 15--21 (1988).

\bibitem[{British Geological Survey}(2012)]{IGRF}
{British Geological Survey}, {The IGRF Model.},
  \url{www.geomag.bgs.ac.uk/gifs/igrf.html} (2012).

\bibitem[{Singh} et~al.(2011)]{Singh2011}
K.~{Singh}, M.~{Mevius}, O.~{Scholten}, et~al., \emph{ArXiv e-prints}  (2011),
  \eprint{1108.5745}.

\end{thebibliography}

\IfFileExists{\jobname.bbl}{}
 {\typeout{}
  \typeout{******************************************}
  \typeout{** Please run "bibtex \jobname" to optain}
  \typeout{** the bibliography and then re-run LaTeX}
  \typeout{** twice to fix the references!}
  \typeout{******************************************}
  \typeout{}
 }

\end{document}